\documentclass[floatfix,showpacs,aps]{revtex4}
\usepackage{amsmath,amssymb}
\usepackage{amsfonts}
\usepackage[latin1]{inputenc}
\usepackage{epsfig}
\usepackage{graphicx}
\usepackage{epsfig}
\usepackage{bm}

\begin{document}

\title{Mean Field Model of Coagulation and Annihilation
Reactions In a Medium of Quenched Traps: Subdiffusion}

\author{I. M. Sokolov$^{1}$, S. B. Yuste$^{2}$, J. J. Ruiz-Lorenzo$^{2}$, and
Katja Lindenberg$^{3}$}
\affiliation{$^{1}$Humboldt University, Newtonstr. 15, D-12489 Berlin,
Germany\\
$^{2}$Departamento de F\'{\i}sica, Universidad de Extremadura, E--6-71
Badajoz, Spain\\
$^{3}$Department of Chemistry and Biochemistry and Institute for
Nonlinear Science, University of California San Diego, 9500 Gilman
Drive, La Jolla, CA 92093-0340, USA}

\begin{abstract}
We present a mean field model for coagulation ($A+A\rightarrow A$) and annihilation ($A+A\rightarrow
0$) reactions on lattices of traps with a distribution of depths reflected in a distribution of mean
escape times.  The escape time from each trap is exponentially distributed about the mean for that
trap, and the distribution of mean escape times is a power law.  Even in the absence of reactions,
the distribution of particles over sites changes with time as particles are caught in ever deeper
traps, that is, the distribution exhibits aging.  Our main goal is to explore whether the reactions
lead to further (time dependent) changes in this distribution.
\end{abstract}


\maketitle

\section{Introduction}

Chemical reactions limited by the motions of the reactants are abundant
in nature, and among the most thoroughly studied are diffusion-limited
reactions. A ubiquitous approach to these systems simply adds the
diffusion and reaction contributions together so as to construct
appropriate reaction-diffusion equations.  It is implicit and even
explicit in these approaches that the diffusive component describes a
motion without a memory, and if one invokes an underlying continuous
time random walk (CTRW) where walkers react when they meet, it is understood that the waiting
time distributions for reactants to remain at one location before moving
on have a finite mean.  The most frequently invoked waiting time in this
scenario is exponential.
It is also understood that in fact many
microscopic models can be subsumed under the same mesoscopic reaction-diffusion
umbrella~\cite{havlin87,benavraham00,blumen86,kang85,toussaint83}.

On the other hand, contrary to the diffusive case,
it is by now fairly clear that different microscopic scenarios of
reactions among subdiffusive species, even simple scenarios, lead to
different mesoscopic
descriptions~\cite{sokolov06,henry06,seki03,sagues08,reichman08}.
For instance, a popular mesoscopic vehicle, the CTRW,
involves waiting time distributions of ensembles of particles undergoing reactions.
The forms of these distributions depend on the underlying microscopic rules
and may vary from one microscopic scenario to another.  It is thus risky to simply assume a form
for these distributions as one would for diffusive particles.  Instead, a more detailed derivation
starting from a set of microscopic rules to arrive at a mesoscopic CTRW description is necessary.
The specific microscopic reaction scenario of interest to us is a lattice whose sites
are occupied by traps of varying depths, that is, the quenched trap scenario.
If independent particles simply walk on this lattice,
their escape time from each trap is exponentially distributed, with the
distribution of trap depths reflected in a distribution of mean escape times.
An average over this distribution leads to a
CTRW model in which all sites have the same waiting time
distribution~\cite{scher,klafter.silbey}. This latter
process is spatially homogeneous and semi-Markovian,
and anomalous diffusion arises if the resulting mean waiting time on each site diverges.
This is the so-called annealed trap scenario.
However, if particles can also react with one another, interesting questions immediately arise.
Of interest to us are the reactions $\mathrm{A} + \mathrm{A} \rightarrow
\mathrm{A}$ (coagulation) and
$\mathrm{A} + \mathrm{A} \rightarrow 0$ (annihilation); 
some uncertainties arising from different microscopic descriptions can be illustrated with
the coagulation reaction.  Suppose there is an $\mathrm{A}$ at a site, and a
second $\mathrm{A}$ arrives.  In the underlying spatially
disordered quenched trap model, it does not matter which particle is ``killed" by the
reaction since the waiting time distribution for particle departure from any site is
exponential. Such a process is often called memoryless because the future of each particle is
determined only by its present state and not its past, and therefore the differences between the two
victims become irrelevant.  However if, as we subsequently do, we implement a mean field assumption
which turns out to lead to a description that is
asymptotically equivalent to a CTRW with diverging mean waiting times,
that is, to an annealed trap model,
it may make a kinetic difference which of the two particles is ``killed" in each reaction.
It is not clear in general how to include such fine distinctions
in a mesoscopic formalism, nor is it clear what sort of rule at the
mesoscopic or mean field level best mimics the behavior of the underlying trapping problem.

Our mean field model shares with other CTRW models with diverging mean waiting times the phenomenon
of ``aging"~\cite{bouchaud2,monthus,barkai,jsp}. Even in non-reactive systems, aging causes the waiting
time distribution itself to change with time as particles settle into sites with ever longer waiting
times (corresponding to particles caught in ever deeper traps
in the underlying system). Aging for non-reactive particles does not occur in a given
finite quenched trap environment since the distribution of particles approaches a Boltzmann
distribution~\cite{barkai2}.  The most intriguing unanswered question that we address
in this work is whether the reactions cause a change in the waiting time distribution.
We anticipate our answer to this latter question: we find that in our mean field model
the reaction does \emph{not} cause an additional change in the waiting time distribution.

We will compare our model predictions with the results of numerical
simulations of the actual distributed trap scenario. We find that the
mean field approach works well for higher dimensions ($d\geq 3$) but not
for low dimensions ($d=1$), a result that agrees with the well-known
differences in subdiffusion exponents predicted by CTRW theories and
those obtained by numerical simulations in the absence of
reactions~\cite{bouchaud}
The concentrations of surviving particles are known to be well reproduced in all dimensions
by invoking relations with the number of distinct sites visited in the asymptotically equivalent
CTRW~\cite{blumen86,ourchapter}.
On the other hand, the mean field formalism, while
restricted to higher dimensions, provides additional direct
insights into the more detailed information contained in the time
dependence of the waiting time distributions.

In Sec.~\ref{ctrwmodel} we describe our mean field model and arrive at a
master equation for the density of particles with given mean
rate for leaving any site in the absence of reactions.  This equation
explicitly shows the effects of aging. In Sec.~\ref{withreaction} we
establish the corresponding master equations in the presence of the
coagulation and annihilation reactions. In Sec.~\ref{solution} we
discuss the solution of the master equation for the time-dependent rate
distribution.  Comparisons with numerical simulation results
for the underlying random trap model are presented in Sec.~\ref{numerical}.
We conclude with a brief summary in Sec.~\ref{conclusions}.

\section{Mean-field approach to a trap model}
\label{ctrwmodel}

To construct a mean field model, we start with a lattice whose sites are traps of varying depths.
Our model is equivalent to that of~\cite{monthus}, but our notation
is suitably modified to facilitate the inclusion of reactions, which they do not consider.
The waiting time for leaving a trap $i$ is exponentially distributed,
$p(t|\tau)=\tau_i^{-1} \exp(-t/\tau_i)$,
where $\tau_i$, the mean sojourn time in the trap, is determined by
the trap's depth via the Kramers (Arrhenius) law.  We further
assume that the distribution of mean waiting times is a power law,
for example,
$p(\tau) = \gamma \tau^{-1-\gamma} \theta (\tau -1)$.
Note that the asymptotics of the pdf $\psi(t)$
of waiting times $t$ upon averaging over this distribution of mean
waiting times is then of power law form even though the waiting time for
each trap is exponentially distributed, i.e.,
\begin{equation}
\psi(t) = \int_0^\infty p(t|\tau) p(\tau) d \tau \sim \gamma \Gamma(1+\gamma) \tau^{-1-\gamma}.
\label{psit}
\end{equation}

The power law distribution of mean waiting times $\tau_i$ leads to
a power law distribution of the ``leaving rates" $\omega_i = 1/\tau_i$ for
departing from a site,
\begin{equation}
p(\omega)= \gamma \omega^{\gamma -1} \theta(1-\omega).
\label{leavingrates}
\end{equation}
Note that all the moments of the distribution of the rates
are finite even if
those of the waiting time distribution are not.

Particles are distributed over sites with different mean waiting times, and as time proceeds, the
distribution of particles over these sites changes even in the absence of reactions because more and
more particles get stuck in deeper and deeper traps.  Our central
question concerns the effects of reactions on this evolving distribution. In particular, we ask
whether the distribution is modified by the reactions.  This is difficult to answer, at least
analytically, without further approximation. The system with distributed traps of varying
depths is spatially inhomogeneous, and if the particle executes the usual nearest
neighbor random walk, the future of any particle moving over this landscape may be
strongly dependent on its particular location.  Furthermore, there
may be strong correlations among subsequent steps (especially in lower
dimensions) because the particle may revisit a previously visited site.
To make progress we implement a mean field approximation designed to provide information about the
evolving distribution of particles over the inhomogeneous landscape.
Specifically, we assume that the particles do not perform a
nearest neighbor random walk 
but instead that each particle is equally likely (probability
$1/N$) to step on any of the $N$ sites of the system.
In other words, the lattice is a complete graph.
Equivalently, as an alternative way of viewing the model, we can think of particles performing
nearest neighbor random walks, but before each step the
mean waiting time associated with the trap to which the particle is about to step is chosen
anew from the distribution $p(\tau)$ or, correspondingly, a
new leaving rate is chosen from $p(\omega)$, independently of the
waiting times chosen in prior steps.  It is also equivalent to  
a nearest neighbor CTRW in a very high-dimensional quenched medium. 
In any case, in the absence of reactions this results in a
space-homogeneous CTRW model with waiting time distribution $\psi(t)$ as given in Eq.~(\ref{psit})
chosen anew at each step.

The assumptions underlying the mean field approach are expected to be
more adequate for random walks that are transient, that is, ones in
which already visited sites are revisited only with a small probability
so that almost all sites reached by the walker are new sites.  This is
the case for random walks in dimensions $d\geq 3$ and, as we show in Sec.~\ref{numerical}, we
do find excellent agreement between the mean field approach and numerical simulations for the
reactions in quenched trap environments for $d\geq 3$.  If the walk
is recurrent, that is, if the same sites are revisited repeatedly,
the approximation may be poor. One-dimensional walks are recurrent, and
$d=2$ is the marginal dimension for this property.

As noted earlier, as time evolves the distribution of particles over the
sites with given mean waiting times (leaving rates) changes even in the absence of the
reactions because more and more particles get stuck in deeper and deeper traps.
We discuss these changes first in the absence and then in the presence of the
reactions.  For this purpose we note that
under the mean field assumption, the master equation
for the probability $a_i(t)$ to be at site $i$ at time $t$ reads
\begin{equation}
\frac{d}{dt}a_i(t) = \frac{1}{N} \sum_{j \neq i} \omega_j a_j(t)
 - \omega_i a_i(t).
\label{ME1}
\end{equation}
Since each site is characterized by its own $\omega$, we can pass
from the occupation probability $a_i$
to the probability $a(\omega,t)$ and introduce the density $n(\omega,t)$
of particles occupying sites with leaving rate between $\omega$ and
$\omega + d \omega$.
One easily deduces that $n(\omega,t)d\omega =a(\omega,t) p(\omega)d\omega$,
where $N p(\omega)d\omega$
is the number of sites with leaving rates between $\omega$ and $\omega + d\omega$.
Note that in a large system
\begin{equation}
\frac{1}{N} \sum_{j \neq i} \omega_j a_j(t) \simeq \frac{1}{N}
\int \omega N p(\omega) a(\omega,t) d\omega=
\int \omega n(\omega,t) d\omega .
\end{equation}
The last integral can be associated with the time-dependent mean rate,
\begin{equation}
{\Omega}(t)=\int \omega n(\omega,t) d\omega.
\label{Mean}
\end{equation}
Multiplying both sides of Eq.~(\ref{ME1}) by $p(\omega)$ leads to
the master equation for $n(\omega,t)$,
\begin{equation}
\frac{d}{dt}n(\omega,t) = {\Omega}(t)p(\omega) - \omega n(\omega,t).
\label{ME2}
\end{equation}
This equation governs the change in the distribution of particles over
jump rates in the absence of reactions.
Note that the total concentration of particles $c = \int n(\omega,t) d\omega$
is constant in time.

\section{Rate distribution with reactions}
\label{withreaction}

Next we explore the effects of reactions on the site occupation probabilities.
We assume that the usual law of mass action is appropriate at the
\emph{local} level.
For the case of the $\mathrm{A}+\mathrm{A} \rightarrow  \mathrm{A}$ reaction
within our mean-field approximation, the change in the occupation
probability at site $i$ is
\begin{equation}
\frac{d}{dt}a_i(t) = [1-a_i(t)]\frac{1}{N} \sum_{j \neq i} \omega_j a_j(t) - \omega_i a_i(t)
\end{equation}
since the number of particles at a site already occupied by
a particle [with probability $a_i(t)$]
does not change upon the arrival of the new particle.
In the case of the $\mathrm{A}+\mathrm{A} \rightarrow  0$ reaction
the number of particles at an occupied site
is reduced by one upon the arrival of a new particle, so that
the corresponding equation reads
\begin{equation}
\frac{d}{dt}a_i(t) = [1-2a_i(t)]\frac{1}{N} \sum_{j \neq i}
\omega_j a_j(t) - \omega_i a_i(t).
\end{equation}
Again we can focus instead on $n(\omega,t)$, but this is no longer
a proper probability density
because the total number of particles is not conserved.
Regrouping  terms one obtains the reaction equations
\begin{equation}
\frac{d}{dt}n(\omega,t) = p(\omega)\int_0^\infty \omega n(\omega,t) d\omega -
\left[\omega+\int_0^\infty \omega n(\omega,t) d\omega \right] n(\omega,t)
\label{RE1}
\end{equation}
for the  $\mathrm{A}+\mathrm{A} \rightarrow  \mathrm{A}$ reaction, and
\begin{equation}
\frac{d}{dt}n(\omega,t) = p(\omega) \int_0^\infty \omega n(\omega,t) d\omega -
\left[\omega+2\int_0^\infty \omega n(\omega,t) d\omega \right] n(\omega,t)
\label{RE2}
\end{equation}
for $\mathrm{A}+\mathrm{A} \rightarrow  0$.
At $t=0$ the particles are homogeneously
distributed over all sites in the system, so the initial condition
for $n$ is $n(\omega,0) = p(\omega)$.
A normalized probability density is obtained by noting that
the overall time-dependent reactant concentration is given by
\begin{equation}
c(t) = \int_0^\infty n(\omega,t) d\omega.
\label{conc}
\end{equation}
The properly normalized probability density of particles
occupying sites with leaving
rate between $\omega$ and $\omega + d\omega$ is then given by
\begin{equation}
p(\omega,t)=\frac{n(\omega, t)}{c(t)}.
\end{equation}
We again introduce the time-dependent mean jumping rate, which is now given
by
\begin{equation}
{\Omega}(t) = c^{-1}(t) \int_0^\infty \omega n(\omega,t) d\omega =
\int_0^\infty \omega p(\omega,t) d\omega,
\label{tdrate}
\end{equation}
and rewrite Eqs.~(\ref{RE1}) and (\ref{RE2}) in the form
\begin{equation}
\frac{d}{dt}n(\omega,t) = c(t){\Omega}(t)p(\omega) -
[\omega+ c(t){\Omega}(t)] n(\omega,t)
\label{RE1_2}
\end{equation}
for the  $\mathrm{A}+\mathrm{A} \rightarrow  \mathrm{A}$ reaction, and
\begin{equation}
\frac{d}{dt}n(\omega,t) = c(t){\Omega}(t)p(\omega) -
[\omega+ 2c(t){\Omega}(t)] n(\omega,t)
\label{RE2_2}
\end{equation}
for the  $\mathrm{A}+\mathrm{A} \rightarrow  0$ reaction.
Integrating these two equations over the $\omega$-domain gives the
classical kinetic equations
\begin{equation}
\frac{d}{dt}c(t) = -\mu {\Omega}(t) c^2(t)
\label{FormK}
\end{equation}
with the stoichiometric coefficient (``molarity'') of the reaction $\mu=1$
for the  $\mathrm{A}+\mathrm{A} \rightarrow  \mathrm{A}$ reaction
and $\mu = 2$ for
$\mathrm{A}+\mathrm{A} \rightarrow 0$.  The mean jump
rate $\Omega(t)$ is thus the time-dependent reaction rate.

The equation for the probability density $p(\omega,t)
= n(\omega,t)c^{-1}(t)$
corresponding to Eqs.~(\ref{RE1_2}) and (\ref{RE2_2}) is then
\begin{equation}
\frac{d}{dt}p(\omega,t)+p(\omega,t) \frac{1}{c(t)}\frac{d}{dt}c(t)
= {\Omega}(t)p(\omega)
- \omega p(\omega,t) -\mu c(t) {\Omega}(t) p(\omega,t).
\end{equation}
With Eq.~(\ref{FormK}) we see that the term with the time derivative
$dc/dt$ on the left side of this equation
and the last term on the right side cancel.
Thus, the final equation for $p(\omega,t)$ does not depend on
the molarity $\mu$ of the
reaction and is the same as the equation for $n(\omega,t)$ in
the absence of reactions,
\begin{equation}
\frac{d}{dt}p(\omega,t) = {\Omega}(t)p(\omega) - \omega p(\omega,t).
\label{EqForP}
\end{equation}
This statement also leads to the remarkable conclusion that within the
mean field approximation adopted in this model
\textit{the reaction does not affect the waiting time
distribution}.  This answers one of our main questions.

\section{Solution for the time-dependent rate distribution}
\label{solution}

The principal question posed earlier, namely, whether the reaction
changes the waiting time distribution,
has been answered in the negative within our mean-field approach.
It is now instructive to obtain an explicit expression for the
time dependent reaction rate. Equation~(\ref{EqForP}) is an
integro-differential equation for
$p(\omega,t)$ since ${\Omega}(t)$ as defined in Eq.~(\ref{tdrate})
depends on $p(\omega,t)$ itself. However, contrary to the equations for
$n(\omega,t)$, this
equation is linear and can best be approached via Laplace transforms.
First we consider the time-Laplace transforms of $p(\omega,t)$
and ${\Omega}(t)$,
\begin{equation}
\tilde{p}(\omega,s)=\int_0^\infty p(\omega,t) e^{-st} dt, \qquad
\tilde{\Omega}(s)=\int_0^\infty {\Omega}(t) e^{-st} dt.
\end{equation}
The transform of Eq.~(\ref{EqForP}) is
\begin{equation}
s\tilde{p}(\omega,s)-p(\omega)= \tilde{\Omega}(s)p(\omega) -\omega \tilde{p}(\omega,s),
\end{equation}
where we have explicitly used the initial condition $p(\omega,0)=p(\omega)$.
The formal ``solution" for $\tilde{p}(\omega,s)$ (with $\tilde{p}$ still
contained in $\tilde{\Omega}$) is
\begin{equation}
\tilde{p}(\omega,s) = [1+\tilde{\Omega}(s)]\frac{p(\omega)}{s+\omega}.
\label{ns}
\end{equation}
Transforming the definition (\ref{tdrate}) of ${\Omega}(t)$ we obtain
\begin{equation}
\tilde{\Omega}(s) = \int_0^\infty \tilde{p}(\omega,s) \omega d\omega,
\end{equation}
so that multiplying both sides of Eq.~(\ref{ns}) by $\omega$ and
integrating, we arrive at a closed algebraic equation for $\tilde{\Omega}(s)$,
\begin{equation}
\tilde{\Omega}(s) = [1+\tilde{\Omega}(s)]\int_0^\infty
\frac{\omega p(\omega)}{s+\omega} d\omega.
\end{equation}
The solution of this equation is
\begin{equation}
\label{Omegas}
\tilde{\Omega}(s) =  \frac{\tilde{I}(s)}{1-\tilde{I}(s)},
\end{equation}
where
\begin{equation}
\tilde{I}(s) \equiv \int_0^\infty \frac{\omega p(\omega)}{s+\omega} d\omega.
\label{IsAsintGral}
\end{equation}

The integral representing $I(s)$ can be evaluated asymptotically for any
long-tailed distribution $p(\tau)$
using the asymptotic method for integrals with weak singularity (see Ch. 1 §4 in~\cite{Fedoryuk}),
as follows. We can rewrite $I(s)$
as
\begin{equation}
I(s) = 1-s\int_0^\infty \frac {p(\omega)}{s+\omega} d\omega = 1
-s\int_0^\varepsilon \frac {p(\omega)}{s+\omega} d\omega
-s\int_\varepsilon^\infty \frac {p(\omega)}{s+\omega} d\omega
\end{equation}
for any $\varepsilon$.  Relevant to the long-time asymptotic behavior is the
small-$s$ behavior of $I(s)$.  As long as $\varepsilon$ is chosen such that $\varepsilon/s \to
\infty$ as $s\to 0$, the third integral on the right hand side is of $O(1)$ in this limit.
Furthermore, if $p(\omega) \sim \gamma \omega^{\gamma-1}$ as $\omega \to 0$ and $\varepsilon \ll 1$,
we can write
\begin{equation}
I(s) \underset {s\to 0}{\sim} 1-s\gamma \int_0^\varepsilon \frac{\omega^{\gamma-1}}{s+\omega}
d\omega -  O(s) \text{\quad with\quad}\varepsilon \ll 1.
\end{equation}
Finally, a change of variables $z=\varepsilon/s$ then allows us to write for $s\to 0$
\begin{equation}
\begin{aligned}
I(s) &\sim 1-\gamma s^\gamma \int_0^{z\to\infty} \frac {z^{\gamma-1}}{1+z} dz -
O(s)\\
& \sim 1-\frac{\gamma \pi}{\sin \pi \gamma} s^\gamma.
\end{aligned}
\label{Isasint}
\end{equation}
For the particular form
$p(\omega)=\gamma \omega^{\gamma-1} \Theta(1-\omega)$ introduced earlier,
one finds the result valid for all $s$ [Abramowitz and Stegun formula 15.3.1]
\begin{equation}
\tilde{I}(s) = \frac{\gamma}{(1+\gamma)s} ~_2F_1(1,1+\gamma,2+\gamma,-1/s),
\end{equation}
which for small $s$ (relevant to long-time asymptotic behavior) leads to the second line of
Eq.~(\ref{Isasint}).
The inverse transform of $\tilde{\Omega}(s)$ then follows
upon application of the Tauberian theorem,
\begin{equation}
{\Omega}(t) \sim \frac{\sin \pi \gamma}{\gamma
\pi \Gamma(\gamma)} t^{\gamma -1}.
\end{equation}
We have thus provided a mesoscopic theoretical foundation for the
widely accepted result that the reaction rate decays with time.

The time dependent reaction rate can now be inserted in the rate
equation (\ref{FormK}) to solve for the concentration as a function of
time.  At long times we find the explicit result
\begin{equation}
c(t) \sim \frac{\pi\gamma \Gamma(1+\gamma)}{\mu \sin \pi \gamma} t^{-\gamma}.
\label{concentration}
\end{equation}
We can in fact calculate the full rate distribution by integrating
Eq.~(\ref{EqForP}). The result up to quadrature is
\begin{equation}
p(\omega,t) = p(\omega)e^{-\omega t}\left[ \int_0^t dt' e^{\omega t'}
\Omega(t') +1\right].
\label{full}
\end{equation}
From this, we can extract the $\omega$ dependence for short times $\omega t \ll 1$ as $p(\omega,t)
\sim p(\omega)$ and for long times $\omega t \gg 1$ as $p(\omega,t)\sim p(\omega)/\omega$.

Finally, we note the interesting connections between these results and
the number $S(t)$ of distinct sites visited up to time $t$ by a
particle in a CTRW~\cite{YKKbriefreport}.
We can connect this quantity to the average
reaction rate $\Omega(t)$ for $d\ge 3$ as follows.  Since at each step of the process a
particle mostly visits a new site in the system
(the random walk is ``transient" or ``non-recurrent"), the reaction
rate can be approximated by the time derivative
of the number of newly visited sites, $\Omega(t)=dS/dt$,
and we have \cite{YKKbriefreport}
\begin{equation}
\tilde{S}(s) \sim \frac{1}{s}\frac{1-R}{1-\tilde{\psi}(s)}, \quad s\to 0
\label{Ss}
\end{equation}
where $R$ is the probability of return to the origin ($R=0.3405\ldots$ for a simple cubic lattice).
Note that the small-$s$ behavior of $\tilde{\psi}(s)$ corresponding
to the asymptotic behavior of $\psi(t)$ in Eq.~(\ref{psit}) is
\begin{equation}
\tilde{\psi}(s)\sim 1-\frac{\gamma \pi}{\sin \pi \gamma} s^\gamma \equiv 1-(s/\lambda)^\gamma
\end{equation}
Inserting this expression into~\eqref{Ss},  inserting~\eqref{Isasint} into~\eqref{Omegas}, and comparing
the resulting expressions, one sees
that $S(s)\sim (1-R)\tilde{\Omega}(s)/s$ for small $s$, so that
$\Omega(t)\sim (1-R)dS/dt$ for large $t$.
Although the interpretation of the reaction rate in terms of
the number of distinct sites visited is quite standard, the fact that
the broad distribution of trapping times does not introduce
any additional fluctuation effects into the kinetics is not at all trivial.

A further connection with the distinct number of sites visited in a CTRW
occurs for the concentration of surviving reactant, namely,
$c(t)\sim 1/S(t)$, a connection that holds
not only for $d>2$ but also for $d\leq 2$, i.e., even
when the random walk is recurrent~\cite{blumen86,ourchapter}.
In \cite{YKKbriefreport} we explicitly obtained the $3d$ result
$S(t) \sim [(1-R)/\Gamma(1+\gamma)] (\lambda t)^\gamma$.  In one dimension
$S(t)\sim [\sqrt{2}/\Gamma(1+\gamma/2)] (\lambda t)^{\gamma/2}$.
We will test the proposition that $c(t) \sim 1/S(t)$ in the
next section along with results of the theory we have developed above.

\section{Numerical results}
\label{numerical}

Our discussion supports the expectation that the results of the mean
field model approximate those of the underlying random depth trap system in
$3d$ (but not in $1d$).  Even in $3d$, where walks are transient, it is
nevertheless the case that there is a finite probability $R$
of return to a previously visited site, e.g. about 1/3 for a simple cubic lattice.
In this section we compare our results with those of numerical
simulations of $3d$ and $1d$ lattices with traps of random mean exit times
as described earlier. The algorithm used in the simulations is described
in some detail in the Appendix.

Figure~\ref{conc3d} shows a collection of results for the concentration
of reactants as
a function of time in $3d$.  The results fall essentially on two straight
lines, except for finite size effects.
The upper set of results is for $\gamma=0.5$ and the lower set for
$\gamma=0.8$.  The concentrations shown are indicated as $c_1(t)$, which
we associate with the coagulation reaction
$\mathrm{A}+\mathrm{A}\rightarrow \mathrm{A}$, and $c_2(t)$ for the
annihilation reaction $\mathrm{A}+\mathrm{A}\rightarrow 0$.
We plot $c_1(t)$ and $2c_2(t)$ to ascertain that
the surviving concentration in the coagulation reaction is twice that of
the annihilation reaction.  The coincidence of the
simulation results for the two cases shows that this is indeed the case.
Simulation results are shown for two sizes of the
$L\times L\times L$ simple cubic lattice with $L=30$ and $L=60$. The
results for the two sizes are the same except for the larger value of
$\gamma$, where we see some deviations from the straight line at very
long times due to finite size effects.  These deviations are more pronounced for the larger $\gamma$
(in the faster walk the ends of the lattices are reached earlier) and for the smaller lattice.
Since this is a log-log plot the straight-line behavior confirms the power law decay of the
concentrations.
Furthermore, we find that the slopes of these lines
are close to $\gamma$.  In particular, we find slopes $0.491(2)$ for
$\gamma=0.5$ and $0.763(2)$ for $\gamma=0.8$ (in both cases for
both reactions).  The agreement in the former case is very good, but somewhat less so
in the latter, where we are not in the fully asymptotic scenario~\cite{footnote}.

\begin{figure}[h!]
\centering
\scalebox{0.99}{\includegraphics[angle=0]{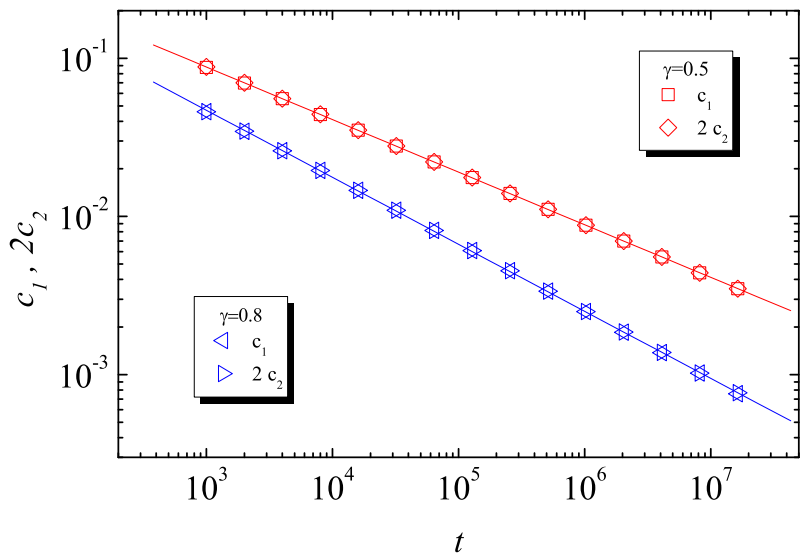}}
\caption{The reactant concentrations $c_1(t)$ and $2c_2(t)$ in one dimension
as functions of time for $\gamma=0.5$ and $\gamma=0.8$.
The symbols denote the results of numerical simulations, and the lines are linear fits to these
results.}
\label{conc1d}
\end{figure}

Figure~\ref{conc1d} shows our simulation results in $1d$, again for
$\gamma=0.5$ and $\gamma=0.8$. The lines in this case are power law fits to the
simulation results.
The size of the lattice is $L=10,000$.  The mean field approach
is not valid here as an approximation
to the underlying trap model, and yet a number of interesting results
are nevertheless worth mentioning. First, we note that the relation
$c_1(t)=2c_2(t)$ still holds.  We do not know whether this indicates
that here again the reactions do not affect the dynamics of the moving
species, but this result indicates that it may be so. Second, we note
that the exponents of the concentration, while quite different from
those of the mean field model,
agree with those associated with the distinct
number of sites visited for the original trap model, namely, that in one
dimension  $c(t) \sim 1/S(t) \sim \left\langle x^2 \right
\rangle ^{-1/2}\sim t^{-\gamma/(1+\gamma)}$ because all sites within the
span of the random walk are visited at least once.
The numerical fits for the simulation results yield
$c_1(t)= 2c_2(t) \sim 0.881(1) t^{-0.3328(1)}$ for $\gamma=0.5$ and
$c_1(t)=2c_2(t) \sim0.959(4) t^{-0.430(1)}$ for $\gamma=0.8$.
The exponents $0.33$ and $0.43$  are in good agreement with
the values of $\gamma/(1+\gamma) = 0.333$ and $0.444$ respectively.

\begin{figure}[h!]
\centering
\scalebox{0.99}{\includegraphics{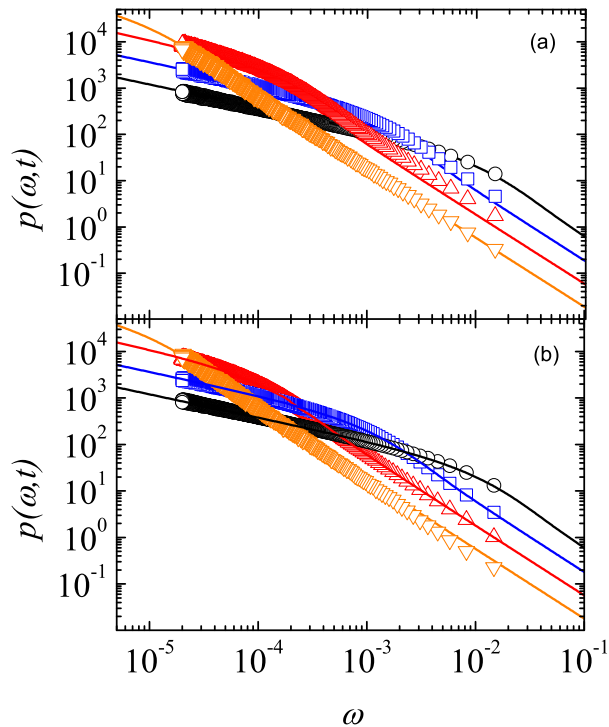}}
\caption{
For all results in this figure, $\gamma=0.5$.
The solid lines are calculated from
the analytic result Eq.~(\ref{full}) and the symbols are the simulation results for lattices
with quenched traps scaled as described in the text.  From high to low on the right side of each
panel the curves are for  $t=10^2$, $10^3$, $10^4$, and $10^5$. 
Top panel: $3d$ lattice; bottom panel: lattice with equally likely jumps to any site, that is, a
complete graph.}
\label{distributions}
\end{figure}

Finally, we compare our analytic mean field predictions in more detail with numerical
simulations of the quenched trap model.  In Fig.~\ref{distributions} we show results for the
distribution $p(\omega,t)$ vs $\omega$.  The solid lines in both panels correspond to the mean field
result~(\ref{full}) for $\gamma=0.5$.  From high to low on the right side of each panel the
curves are for $t=10^2$, $10^3$, $10^4$, and $10^5$.  Equation~(\ref{full}) leads to
$p(\omega,t)\sim p(\omega) \sim \omega^{\gamma-1}$ for $\omega t \ll 1$ and
$p(\omega,t)\sim p(\omega)/\omega \sim \omega^{\gamma-2}$ for $\omega t \gg 1$;
the slopes $-0.5$ and $-1.5$ in the log-log
plot are evident. The symbols show the corresponding simulation results.
The top panel shows the results for a $3d$ quenched trap lattice with
nearest neighbor steps, and the bottom panel those of of a quenched trap lattice in which steps to
any site are equally likely. Our only adjustment in these figures is the normalization, which can
not be obtained properly from the simulations since they do not cover an infinite range of $\omega$s.
We have made this adjustment by  scaling the simulation results so that the quantity $R(t)
=\int_{\omega_{min}}^{\omega_{max}} P(\omega,t) d\omega < 1$ evaluated using the scaled results
agrees with the one obtained by means of the theoretical result~(\ref{full}). Here
$\omega_{min}$ and $\omega_{max}$ define the range of $\omega$s covered by the
simulations. These results confirm that already in $3d$ the mean field theory captures the behavior as
well as does the actual numerical realization of the model. We have also ascertained the agreement
between simulations and mean field theory for other values of the subdiffusive exponent $\gamma$.

\begin{figure}[h!]
\centering
\scalebox{0.99}{\includegraphics[angle=0]{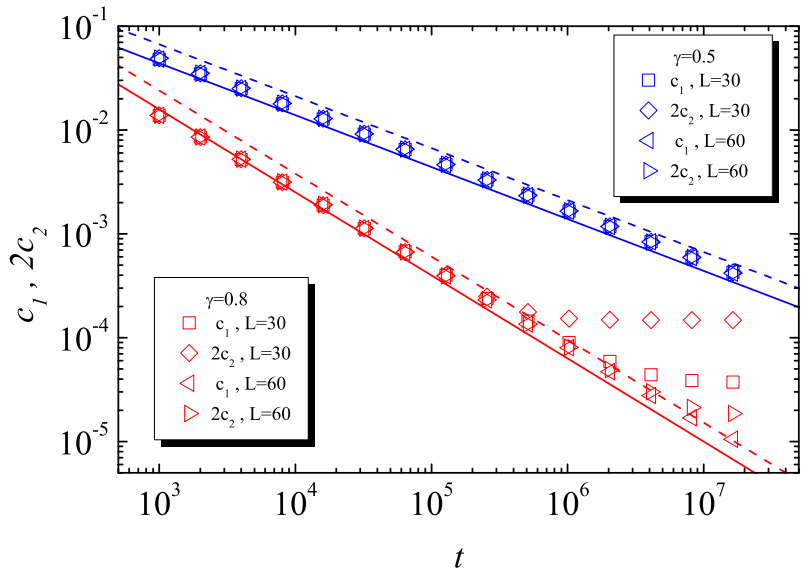}}
\caption{The reactant concentrations $c_1(t)$ and $2c_2(t)$
in three dimensions as functions of time for $\gamma=0.5$
and $\gamma=0.8$ and two lattice sizes.  The symbols denote numerical simulation results.
The solid lines result from the mean field theory, and the dashed lines from the connection with the
number of distinct sites visited.}
\label{conc3d}
\end{figure}

\section{Conclusions}
\label{conclusions}

We have presented a mean field theory of for coagulation ($\mathrm{A} + \mathrm{A} \rightarrow
\mathrm{A}$) and annihilation ($\mathrm{A} + \mathrm{A} \rightarrow 0$) reactions on a lattice whose
sites are occupied by traps of varying depths. The escape times from these traps are distributed
exponentially about a mean time, and the distribution of mean escape times is of power law form.
We calculate reactant concentrations as a function of time as well as the time-dependent
distribution of particles over sites with different escape rates.
The mean field model is designed with the particular goal of studying this evolving distribution and
the effects of the reactions on it, and is expected to do well in dimensions $d\geq 3$ but not
in $1d$.  The noteworthy outcome of the model is
that this distribution is \emph{not changed} by the occurrence of the reactions.
One consequence of this result is that the concentration of surviving reactant in the
$A+A\rightarrow A$ reaction is double that of the $A+A\rightarrow 0$ reactions, that is,
$c_1(t)=2c_2(t)$. While the model does not shed light on the $1d$ case, numerical simulations here
also show that $c_1(t)=2c_2(t)$ and thus a reaction-insensitive aging distribution of particles is
not ruled out.
Numerical simulations in $d = 3$ with quenched traps agree with these predictions.

While the waiting time distribution is unaffected by the reaction
in this mean field model, it is interesting
to speculate about the assumptions that would have to be made in a
\emph{spatially translationally invariant model with long-tailed waiting time distributions}
to arrive at this conclusion in a formulation
that includes an explicit description of the reaction
(we remind the reader that this is not an issue in the underlying
random trap depth model with exponentially distributed waiting times).
The result would seem to be automatically
correct for the $\mathrm{A}+\mathrm{A} \rightarrow 0$ reaction
since every reacting pair involves
two particles that have not been previously involved in a reaction.
However, for the
$\mathrm{A}+\mathrm{A} \rightarrow \mathrm{A}$ reaction the situation
is different.  Here one of the
reaction partners continues its walk and may participate in a
later reaction. It may make a
difference whether the survivor is the $\mathrm{A}$ that arrived
at the site immediately before the
reaction (``no kill" scenario), or the $\mathrm{A}$ that was there already
(``kill" scenario), or a choice of one or the other
according to some probability.  While it might be tempting to assume
that a random choice of one or the
other (e.g. with equal probability) would lead to the mean
field result obtained above, it is not immediately evident
that this choice provides exactly the correct compensatory effect.
Interestingly, in our numerical simulations we find no statistically significant difference between
results obtained in the two scenarios.

\section*{Acknowledgments}

The research of S.B.Y. has been supported by the Ministerio de Educaci\'on y Ciencia (Spain) through grant
No. FIS2007-60977 (partially financed by FEDER funds). I.S. thanks the DFG for support through the
joint collaborative program SFB 555.  K.L. gratefully acknowledges the support of the National
Science Foundation under Grant No. PHY-0354937.

\appendix*
\section{Description of the numerical algorithm}

In this appendix we describe the steps used in our numerical algorithm.
\begin{enumerate}
\item We generate a $1d$ or $3d$ lattice.
\item
We generate a mean escape time $\tau_i$ to be associated with each lattice site
$i$ using a given probability distribution. In this
paper we use the probability distribution $p(\tau) = \gamma \tau^{-1-\gamma} \theta (\tau -1)$. The
times $\tau_i$ are fixed throughout the entire simulation.

\item
We distributed particles on the lattice sites with concentration $c$ so that the probability that
any particular site is initially occupied is $c$. There is at most one particle per site.
\item
The dynamics then proceeds as follows:
\begin{enumerate}
\item
\label{4.1}
We generate the waiting times for each particle's next jump using an exponential distribution about
a mean escape time $\tau_i$ associated with each site $i$.
\item
\label{4.2}
We choose the particle with the smallest waiting time.  This particle jumps to one of its nearest
neighbors (two in $1d$, six in $3d$) with equal probability.
\item
\label{4.3}
If the destination site is empty, we update the waiting time for the arriving particle (we simply
add time to the old waiting time a new time obtained using~\ref{4.1} and we repeat the
process~\ref{4.2}.
\item
If the destination site is occupied, the particles annihilate (in the $A+A\rightarrow 0$) or
coagulate (in the case $A+A\rightarrow A$).  We then repeat the process by returning to~\ref{4.2}.
In the $A+A\rightarrow A$ case we need to specify which of the two particles is the victim 
in the coagulation process. In the ``kill"
scenario the arriving particle (the particle that just performed the jump) annihilates the particle
that was already there, and the waiting time of the arriving particle is updated as in~\ref{4.3},
that is, as if the destination site were empty. In the ``no kill" scenario the arriving particle is
annihilated, and the waiting time of the surviving particle remains unchanged.  One could also
implement a combination of these rules with some probability weighting.  In any case, in our
simulations interestingly we have observed no statistically significant difference between the
results obtained with "kill" and with "no kill" rules.
\end{enumerate}
\end{enumerate}


\begin{thebibliography}{99}

\bibitem{havlin87}
S. Havlin and D. ben-Avraham, Adv. Phys. {\bf 36}, 695 (1987).

\bibitem{benavraham00}
D. ben-Avraham and S. Havlin, \emph{Diffusion and Reactions in Fractals
and Disordered Systems} (Cambridge University Press, Cambridge, 2000).

\bibitem{blumen86}
A. Blumen, J. Klafter and G. Zumofen, in \emph{Optical Spectroscopy of
Gladdes}, edited by J. Zschokke (Reidel, Dordrecht, 1986).

\bibitem{kang85}
K. Kang and S. Redner, Phys. Rev. A {\bf 32}, 435 (1985).

\bibitem{toussaint83}
D. Toussaint and F. Wilczek, J. Chem. Phys. {\bf 78}, 2642 (1983).

\bibitem{sokolov06}
I. M. Sokolov, M. G W. Schmidt, and F. Sagu\'es, Phys. Rev. E {\bf 73},
031102 (2006).

\bibitem{henry06}
B. I. Henry, T. A. M. Langlands, and S. L. Wearne, Phys. Rev. E {\bf
74}, 031116 (2006).

\bibitem{seki03}
K. Seki, M. Wojcik, and M. Tachiya, J. Chem. Phys. {\bf 119}, 2165
(2003).

\bibitem{sagues08}
F. Sagu\'es, V. P. Shkilev, and I. M. Sokolov, Phys. Rev. E {\bf 77},
032102 (2008).

\bibitem{reichman08}
J. D. Eaves and D. R. Reichman, J. Phys. Chem. B {\bf 112}, 4283 (2008).

\bibitem{scher}
H. Scher and E. W. Montroll, Phys. Rev. B {\bf 12}, 2455 (1975).

\bibitem{klafter.silbey}
J. Klafter and R. J. Silbey, Surface Science {\bf 92}, 393 (1980).

\bibitem{bouchaud2}
J. P. Bouchaud, J. Phys. I France {\bf 2}, 1705 (1992).

\bibitem{monthus}
C. Monthus and J-P. Bouchaud, J. Phys. A: Math. Gen. {\bf 29}, 3847 (1996).

\bibitem{barkai}
E. Barkai, Phys. Rev. Lett. {\bf 90}, 104101 (2003).

\bibitem{jsp}
V. Yu. Zaburdaev, J. Stat. Phys. {\bf 133}, 159 (2008).

\bibitem{barkai2}
S. Burov and E. Barkai, Phys. Rev. Lett. {\bf 98}, 250601 (2007).

\bibitem{bouchaud}
J. P. Bouchaud and A. Georges, Phys. Rep. {\bf 195}, 127 (1990).

\bibitem{YKKbriefreport}
S. B. Yuste, J. Klafter, K. Lindenberg, Phys. Rev. E 77, 032101 (2008)

\bibitem{ourchapter}
S. B. Yuste, K. Lindenberg, and J. J. Ruiz-Lorenzo, in \emph{Anomalous Transport: Foundations and
Applications}, ed. by R. Klages, G Radons, and I. M. Sokolov (Wiley-VCH, Berlin, 2008).

\bibitem{Fedoryuk}
M. V. Fedoryuk, Asymptotics: Integrals and Series, Nauka,
Moscow, 1987 (in Russian).

\bibitem{footnote}
We have computed the prefactor in the power law fit, but it is very difficult to obtain reliable
values for it since a small variation in the exponent, e.g., from 0.491 to 0.5, results in a
considerable change in the prefactor, e.g. from 1.48(2) to 1.683(2).  In any case, since we have
obtained the same exponent for both reactions, the prefactors are fully compatible (within
statistical error) with the hypothesis $c_1(t)=2 c_2(t)$ for both values of $\gamma$.


\end{thebibliography}
\end{document}